# Quantum analysis of the nondegenerate optical parametric oscillator with injected signal

B. Coutinho dos Santos, K. Dechoum, A. Z. Khoury, and L. F. da Silva

*Instituto de Física da Universidade Federal Fluminense,*

*Boa Viagem 24210-340, Niterói-RJ, Brazil.*

M. K. Olsen

*ARC Centre of Excellence for Quantum-Atom Optics, School of Physical*

*Sciences, University of Queensland, Brisbane, Qld 4072, Australia.*

(Dated: July 8, 2005)

## Abstract

In this paper we study the nondegenerate optical parametric oscillator with injected signal, both analytically and numerically. We develop a perturbation approach which allows us to find approximate analytical solutions, starting from the full equations of motion in the positive P-representation. We demonstrate the regimes of validity of our approximations via comparison with the full stochastic results. We find that, with reasonably low levels of injected signal, the system allows for demonstrations of quantum entanglement and the Einstein-Podolsky-Rosen paradox. In contrast to the normal optical parametric oscillator operating below threshold, these features are demonstrated with relatively intense fields.

## I. INTRODUCTION

Intracavity parametric down-conversion is a relatively simple nonlinear optical process which can exhibit nonclassical behavior and allow for experimental tests of quantum mechanics. In the simplest case, that of degenerate down-conversion, a pump field at frequency $\omega_c$ produces, via interaction with a nonlinear medium, a field at half this frequency. In the nondegenerate parametric oscillator (NOPO), considered in this article, two distinguishable down-converted fields are produced, either with orthogonal polarizations, or with different frequencies, $\omega_a$ and $\omega_b$, such that $\omega_c = \omega_a + \omega_b$. This latter case, with a coherent signal injected into the cavity at the frequency $\omega_b$, allows for increased conversion efficiency over the uninjected case, as well as a good degree of tunability. The injection of the signal field can be used, for example, to choose the frequency of a laser. Rather than producing an output field at either half or twice the frequency of the pump, an injected signal then allows for the production of a macroscopic field at the difference of the signal and pump frequencies, therefore making the system more versatile. As it has been shown that measurements using light are not limited by the Rayleigh criterion, but by the inherent noise of the field [1–3], with amplitude squeezed fields allowing for higher sensitivity, parametric processes with injected signal may be suitable for the production of nonclassical light of a given frequency for measurement and spectroscopic purposes [4].

A number of theoretical analysis of the NOPO have been published over recent years. Among these, an early analysis by Björk and Yamamoto raised the possibility of the production of photon number states by using idler measurement feedback [5]. In a subsequent article the same authors investigated quadrature phase fluctuations in the same system [6], finding fluctuation levels below those of a coherent state. The authors discussed possible applications in optical communications and gravity wave detection and predicted that the suppression of fluctuations would be greatest close to the threshold pumping power. Reid and Drummond investigated the correlations in the NOPO both above [7] and below threshold [8]. In the above threshold case, they studied the effects of phase diffusion in the signal and idler modes, beginning with the positive-P representation equations of motion for the interacting fields [9]. Changing to intensity and phase variables, they were able to show that output quadratures could be chosen which exhibited fluctuations below the coherent state level and also Einstein-Podolsky-Rosen (EPR) type correlations. In the below thresh-old case, a standard linearized calculation was sufficient to obtain similar correlations. Su *et al.* [10] investigated the utility of the NOPO with polarization degeneracy and an optical system with four mirrors for quantum nondemolition measurements of the intensity difference of the two low frequency modes. In the limit of a rapidly decaying pump mode, Kheruntsyan and Petrosyan were able to calculate exactly the steady-state Wigner function for the NOPO, showing clearly the threshold behavior and the phase diffusion above this level of pumping [11].

There are a smaller number of published theoretical investigations of the NOPO with injected signal (INOPO), but we can begin with a linearized analysis by Wong [12], which predicted that the amount of single-beam squeezing above threshold would be greater than without injection. Although not explicitly stated in the article, this is due to the fact that an injected signal stabilizes the phase diffusion considered by Reid and Drummond and also by Kheruntsyan and Petrosyan in the uninjected case. Harrison and Walls [13] studied possible quantum nondemolition measurements of the intensity difference of the two down-converted fields, considering injected signals at both the down-converted frequencies. Zhang *et al.* [14] used a linear fluctuation analysis to study the frequency degenerate but orthogonally polarized output fields of a pumped intracavity type II crystal. Considering injected signals in both the down-converted fields, they found noise suppression in both combined quadratures and in the intensity difference of these two fields.

As far as noise reduction is concerned, experimental attention has been focused on the polarization degenerate NOPO. In an early experiment, noise reduction of 30% below the vacuum level in the intensity difference was measured in the two polarization nondegenerate outputs of an uninjected NOPO operating above threshold [15]. Noise reduction in two-mode combined quadratures and the intensity difference of polarization nondegenerate outputs was measured by Peng *et al.* [16], again operating above threshold and without injection. Zhang *et al.* [17] used a polarization nondegenerate OPO with injected seeds at both polarizations to measure squeezing in a combined quadrature and hence infer an EPR correlation [18] between spatially separated outputs. The production of quadrature entangled light and EPR states has been dealt with theoretically and experimentally in an NOPO operating below threshold, theoretically by using a linearized approximation with the intracavity pump field treated classically [19]. Noise reduction in the difference of the output intensities in both the degenerate and nondegenerate cases without injection have been calculated and measured for input powers up to 14 times the threshold value [20]. Recently, Guo *et al.* reported the design and construction of a compact and portable device using a polarization nondegenerate NOPO with signal injection which can generate nonclassical light for over an hour [21].

There have also been a number of experimental and theoretical investigations of the tunability which the INOPO allows in laser operation, without considering the quantum properties of the output fields. An early demonstration of the ability to select a particular operating wavelength was reported by Boczar and Scharpf [22], who achieved pulsed laser operation at 486.1 nm with a pump wavelength of 354.7 nm and injection at 1312.2 nm. Smith *et al.* [23] developed a classical model of a pulsed INOPO operating inside a ring cavity and compared this to experimental results, finding reasonable agreement with their predictions for spatial beam quality, spectral performance and beam profiles. The ability to efficiently generate tunable frequencies in the mid infrared via injection seeding was demonstrated by Haidar and Ito [24], who also showed a noticeable narrowing of the signal linewidth due to the injection. Bapna and Dasgupta used a seed wave from a dye laser to increase the power and tune the frequency of the output of an INOPO operating in the pulsed regime [25]. What these investigations have shown is that injection can increase the conversion efficiency and stability as well as allowing a large degree of tunability, but they have not investigated the quantum properties of the resulting fields.

In this work our main interests are the effect of the injected signal on the quantum properties of the outputs and the change in threshold behavior as a result of the injection. To this end we will expand the full equations of motion in terms of a perturbation parameter. The zeroth order of this expansion describes the classical behavior while the first order may be thought of as describing linear quantum noise. We will first examine the classical properties of the system, finding approximate analytical solutions and demonstrating their regimes of validity. These will show that the signal and idler fields can be relatively intense at pumping values well below the standard uninjected threshold. We will then solve the first-order equations, the solutions of which allow us to calculate spectral variances outside the cavity. We compare the results found in this manner with the predictions of stochastic integration, finding good agreement. Unlike the analysis of Ref. [7], we may use the normal amplitude variables as the injected signal provides a phase reference which prevents the phase diffusion found in that work. Finally, we will look at the utility of our injected system for the production of entangled states and a demonstration of the EPR correlations, finding that these are now possible with macroscopically intense fields.

## II. HAMILTONIAN AND STOCHASTIC EQUATIONS

The system we consider consists of three modes of the electromagnetic field which are coupled by a nonlinear crystal which is held inside an optical cavity. The three modes have frequencies $\omega_0$, $\omega_1$ and $\omega_2$, where, by energy conservation, $\omega_0 = \omega_1 + \omega_2$. We will consider the case of a strong external driving field, $\mathcal{E}_0$, at frequency $\omega_0$, and a generally much weaker injected field, $\mathcal{E}_1$, at frequency $\omega_1$. There is no input field at frequency $\omega_2$. The intracavity fields at frequency $\omega_j$ are described by the bosonic operators $\hat{a}_0$, $\hat{a}_1$ and $\hat{a}_2$. Following the usual terminology, we shall call the fields represented by these operators the pump, signal and idler, respectively. Each field is damped via the cavity output mirror, interacting with the reservoir fields, denoted by the bath operators $\hat{\Gamma}_j$. The effective second-order nonlinearity of the crystal is represented by the constant $\chi$.

The Heisenberg picture Hamiltonian which describes this system can be written as [26, 27]

$$\begin{aligned}
\hat{H} &= \sum_{i=0}^{2} \hbar\omega_i \hat{a}_i^\dagger \hat{a}_i + i\hbar\chi \left( \hat{a}_1^\dagger \hat{a}_2^\dagger \hat{a}_0 - \hat{a}_1 \hat{a}_2 \hat{a}_0^\dagger \right) + i\hbar \left( \mathcal{E}_0 e^{-i\omega_0 t} \hat{a}_0^\dagger - \mathcal{E}_0^* e^{i\omega_0 t} \hat{a}_0 \right) \\
&+ i\hbar \left( \mathcal{E}_1 e^{-i\omega_1 t} \hat{a}_1^\dagger - \mathcal{E}_1^* e^{i\omega_1 t} \hat{a}_1 \right) + \sum_{i=0}^{2} \left( \hat{a}_i \hat{\Gamma}_i^\dagger + \hat{a}_i^\dagger \hat{\Gamma}_i \right).
\end{aligned} \quad (1)$$

Although exact Heisenberg equations of motion can be found from this Hamiltonian, it is, at the very least, extremely difficult to solve nonlinear operator equations. We will therefore develop stochastic equations of motion in the positive-P representation, which in principle gives access to any normally-ordered operator expectation values which we may wish to calculate. To find the appropriate equations, we proceed via the master and Fokker-Planck equations. Using the standard techniques for elimination of the baths [28], we find the zero-temperature master equation for the reduced density operator.

The master equation may be mapped onto a Fokker-Planck equation [29] for the positive-P pseudoprobability distribution of the six independent complex variables, $(\alpha_0, \alpha_1, \alpha_2, \alpha_0^+, \alpha_1^+, \alpha_2^+)$, whose correlations correspond to those of the normally ordered operators. Using the standard procedure we can derive the following Itô stochastic differential equations [30] in a rotating frame,

$$d\alpha_0 = \left( \mathcal{E}_0 - \gamma_0 \alpha_0 - \chi \alpha_1 \alpha_2 \right) dt,$$

$$\begin{aligned}
d\alpha_0^+ &= \left(\mathcal{E}_0^* - \gamma_0 \alpha_0^+ - \chi \alpha_1^+ \alpha_2^+\right) dt, \\
d\alpha_1 &= \left(\mathcal{E}_1 - \gamma_1 \alpha_1 + \chi \alpha_2^+ \alpha_0\right) dt + (\chi \alpha_0)^{1/2} dW_1, \\
d\alpha_1^+ &= \left(\mathcal{E}_1^* - \gamma_1 \alpha_1^+ + \chi \alpha_2 \alpha_0^+\right) dt + \left(\chi \alpha_0^+\right)^{1/2} dW_1^+, \\
d\alpha_2 &= \left(-\gamma_2 \alpha_2 + \chi \alpha_1^+ \alpha_0\right) dt + (\chi \alpha_0)^{1/2} dW_2, \\
d\alpha_2^+ &= \left(-\gamma_2 \alpha_2^+ + \chi \alpha_1 \alpha_0^+\right) dt + \left(\chi \alpha_0^+\right)^{1/2} dW_2^+,
\end{aligned} \qquad (2)$$

where $\gamma_i$ ($i = 0, 1, 2$) now represent the cavity damping rates at each frequency. In the above, the complex Gaussian noise terms are defined by the relations

$$\begin{aligned}
\overline{dW_1} &= \overline{dW_2} = 0, \\
\overline{dW_1 dW_2} &= \overline{dW_1^+ dW_2^+} = dt.
\end{aligned} \qquad (3)$$

Note here that we are considering that the crystal is perfectly phase matched for the process of down conversion with $\omega_0 = \omega_1 + \omega_2$.

Without loss of generality, we may consider the pump as a real field, $\mathcal{E}_0 = \mathcal{E}_0^* = E_0$, and the signal injection with a relative phase shift $\phi$ so that $\mathcal{E}_1 = E_1 e^{i\phi}$. We therefore find it useful to define the field quadratures

$$X_k = \left(e^{-i\theta_k} \alpha_k + e^{i\theta_k} \alpha_k^+\right), \qquad Y_k = \frac{1}{i}\left(e^{-i\theta_k} \alpha_k - e^{i\theta_k} \alpha_k^+\right), \qquad (4)$$

with $\theta_0 = 0$, $\theta_1 = \phi$ and $\theta_2 = -\phi$. We note here that, with these definitions, the pump and signal are both real in what follows.

For simplicity we will set $\gamma_1 = \gamma_2 = \gamma$, $\gamma_r = \gamma_0/\gamma$, and introduce a scaling parameter,

$$g = \frac{\chi}{\gamma \sqrt{2\gamma_r}}, \qquad (5)$$

in the stochastic equations to make these amenable to perturbation theory [31]. We also introduce a scaled time $\tau = \gamma t$, and the scaled quadratures

$$\begin{aligned}
x_0 &= g\sqrt{2\gamma_r} X_0, & y_0 &= g\sqrt{2\gamma_r} Y_0, \\
x_1 &= g X_1, & y_1 &= g Y_1, \\
x_2 &= g X_2, & y_2 &= g Y_2.
\end{aligned} \qquad (6)$$

In these new variables, the stochastic equations for the quadratures become

$$dx_0 = -\gamma_r \left[x_0 - 2\mu_0 + (x_1 x_2 - y_1 y_2)\right] d\tau,$$

$$dy_0 = -\gamma_r \left[y_0 + (x_1 y_2 + y_1 x_2)\right] d\tau,$$

$$dx_1 = \left[-x_1 + 2\mu_1 + \frac{1}{2}(x_0 x_2 + y_0 y_2)\right] d\tau + \frac{g}{\sqrt{2}} \left[\sqrt{x_0 + iy_0}\, dw_1 + \sqrt{x_0 - iy_0}\, dw_1^+\right],$$

$$dy_1 = \left[-y_1 + \frac{1}{2}(x_2 y_0 - y_2 x_0)\right] d\tau - i\frac{g}{\sqrt{2}} \left[\sqrt{x_0 + iy_0}\, dw_1 - \sqrt{x_0 - iy_0}\, dw_1^+\right],$$

$$dx_2 = \left[-x_2 + \frac{1}{2}(x_0 x_1 + y_0 y_1)\right] d\tau + \frac{g}{\sqrt{2}} \left[\sqrt{x_0 + iy_0}\, dw_2 + \sqrt{x_0 - iy_0}\, dw_2^+\right],$$

$$dy_2 = \left[-y_2 + \frac{1}{2}(x_1 y_0 - y_1 x_0)\right] d\tau - i\frac{g}{\sqrt{2}} \left[\sqrt{x_0 + iy_0}\, dw_2 - \sqrt{x_0 - iy_0}\, dw_2^+\right], \quad (7)$$

where $\mu_0 = E_0 \chi / \gamma \gamma_0$, $\mu_1 = E_1 \chi / \gamma \sqrt{2\gamma \gamma_0}$, $dw_1 = e^{-i\phi} \sqrt{2\gamma}\, dW_1$, $dw_1^+ = e^{i\phi} \sqrt{2\gamma}\, dW_1^+$, $dw_2 = e^{i\phi} \sqrt{2\gamma}\, dW_2$, and $dw_2^+ = e^{-i\phi} \sqrt{2\gamma}\, dW_2^+$. We note here that, in the normal case without signal injection, $\mu_0 = 1$ would indicate the threshold for oscillation. Although there is no longer a true threshold once an injected signal is present, we will continue to use this well-known terminology.

## III. STEADY-STATE MEAN-VALUE SOLUTIONS

Although Eqs. 7 describe the full quantum dynamical behavior of the system, they are very difficult to solve except by numerical simulation. While this is a very powerful technique, sometimes we can gain useful physical insights via approximate analytical solutions. In the present case we find it very useful to examine the classical behavior of the system, solving for the steady-state of Eqs. 7 with the noise terms removed. This procedure leads us to the expressions

$$0 = x_{0s} - 2\mu_0 + (x_{1s} x_{2s} - y_{1s} y_{2s}),$$

$$0 = y_{0s} + (x_{1s} y_{2s} + y_{1s} x_{2s}),$$

$$0 = -x_{1s} + 2\mu_1 + \frac{1}{2}(x_{0s} x_{2s} + y_{0s} y_{2s}),$$

$$0 = -y_{1s} + \frac{1}{2}(x_{2s} y_{0s} - y_{2s} x_{0s}),$$

$$0 = -x_{2s} + \frac{1}{2}(x_{0s} x_{1s} + y_{0s} y_{1s}),$$

$$0 = -y_{2s} + \frac{1}{2}(x_{1s} y_{0s} - y_{1s} x_{0s}), \quad (8)$$

for the quadratures in the steady-state. The solution for the $y$ quadratures is $y_{0s} = y_{1s} = y_{2s} = 0$. Solving the above system for $x_{1s}$ and $x_{2s}$ in terms of $x_{0s}$ we obtain

$$x_{1s} = \frac{8\mu_1}{4 - x_{0s}^2},$$

$$x_{2s} = \frac{4\mu_1 x_{0s}}{4 - x_{0s}^2}. \tag{9}$$

In appendix A we find a set of approximate solutions for $x_{0s}$ valid below,

$$x_{0s} = 2\mu_0 \left[1 - \frac{2\mu_1^2}{(1 - \mu_0^2)^2 + 2\mu_1^2}\right], \tag{10}$$

at,

$$x_{0s} = 2 - (2\mu_1)^{2/3}, \tag{11}$$

and above threshold ($\mu_0 > 1$)

$$x_{0s} = 2 - \frac{\mu_1^2}{2(\mu_0 - 1)} \left[\sqrt{1 + \frac{8(\mu_0 - 1)}{\mu_1^2}} - 1\right]. \tag{12}$$

In Fig. 1a we plot the steady state intracavity pump quadrature amplitude, obtained numerically as a function of the pump input $\mu_0$ for a fixed injection $\mu_1$. Instead of the abrupt transition at threshold, well-known from classical analysis of the the uninjected OPO, the steady state amplitude quadrature asymptotically approaches the threshold value of $x_{0s} = 2$. The solutions obtained from the linear and quadratic approximations are also shown, demonstrating a good agreement with the numerical result well below and well above threshold respectively.

In Fig. 1b we compare the approximate quartic solution at threshold, given by $\mu_0 = 1$, with the full numerical solution for $x_{0s}$ as the injected signal, given by $\mu_1$, increases. We find excellent agreement for injected signal amplitudes up to around a quarter that of the pump.

## IV. PERTURBATION APPROACH FOR QUANTUM FLUCTUATIONS

In this section we will go a step further by using a perturbation expansion of the positive-P representation Eqs. 7. This allows us to include quantum effects in a systematic fashion [32]. We first introduce a formal perturbation expansion in powers of the parameter $g$,

$$x_k = \sum_{n=0}^{\infty} g^n x_k^{(n)},$$

$$y_k = \sum_{n=0}^{\infty} g^n y_k^{(n)}. \tag{13}$$

The series expansion written in this way has the property that the zeroth order term corresponds to the classical field of order 1 in the unscaled quadrature, while the first order

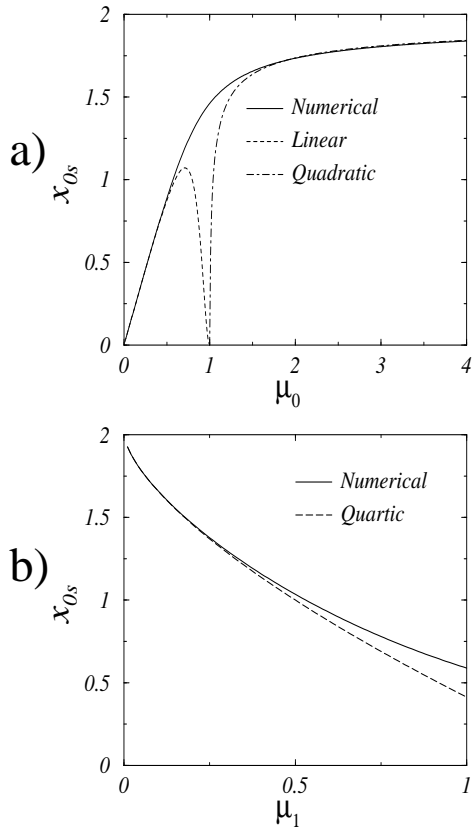

FIG. 1: a) Dimensionless steady state amplitude quadrature $x_{0s}$ as a function of the dimensionless pump parameter $\mu_0$. Linear (dashed line) and quadratic (dot-dashed line) approximations are compared to the numerical calculation (solid line). The dimensionless injection parameter is $\mu_1 = 0.2$. b) Dimensionless steady state amplitude quadrature $x_{0s}$ as a function of the dimensionless injection parameter $\mu_1$. The quartic approximation (dashed line) is compared to the numerical calculation (solid line). The dimensionless pump parameter is $\mu_0 = 1$

.

term is related to quantum fluctuations of order $g$, and the higher order terms correspond to nonlinear corrections to the quantum fluctuations of order $g^2$ and greater. The stochastic equations are then solved by the technique of matching powers of $g$ in the corresponding time evolution equations. This technique can be employed diagrammatically, and so can be termed the "stochastic diagram" method [31] (For related approaches using diagrams and Green's functions, see Refs. [33, 34]).

The zeroth order terms correspond to the classical nonlinear equations of motion for the interacting fields, whose steady state solutions were given in section III. The first order set of equations is then

$$dx_0^{(1)} = -\gamma_r \left[ x_0^{(1)} + x_{1s} x_2^{(1)} + x_{2s} x_1^{(1)} \right] d\tau,$$

$$dy_0^{(1)} = -\gamma_r \left[ y_0^{(1)} + x_{1s} y_2^{(1)} + x_{2s} y_1^{(1)} \right] d\tau,$$

$$dx_1^{(1)} = \left[ -x_1^{(1)} + \frac{1}{2} x_{0s} x_2^{(1)} + \frac{1}{2} x_{2s} x_0^{(1)} \right] d\tau + \sqrt{x_{0s}} dw_{x1},$$

$$dy_1^{(1)} = \left[ -y_1^{(1)} + \frac{1}{2} x_{2s} y_0^{(1)} - \frac{1}{2} x_{0s} y_2^{(1)} \right] d\tau - i\sqrt{x_{0s}} dw_{y1},$$

$$dx_2^{(1)} = \left[ -x_2^{(1)} + \frac{1}{2} x_{0s} x_1^{(1)} + \frac{1}{2} x_{1s} x_0^{(1)} \right] d\tau + \sqrt{x_{0s}} dw_{x2},$$

$$dy_2^{(1)} = \left[ -y_2^{(1)} + \frac{1}{2} x_{1s} y_0^{(1)} - \frac{1}{2} x_{0s} y_1^{(1)} \right] d\tau - i\sqrt{x_{0s}} dw_{y2}. \tag{14}$$

Note that we have defined new Wiener increments as $dw_{x1(y1)}(\tau) = (dw_1(\tau) \pm dw_1^+(\tau))/\sqrt{2}$ and $dw_{x2(y2)}(\tau) = (dw_2(\tau) \pm dw_2^+(\tau))/\sqrt{2}$, with the correlations

$$\langle dw_{x1} dw_{x2} \rangle = d\tau,$$
$$\langle dw_{y1} dw_{y2} \rangle = d\tau, \tag{15}$$

and with all other correlations vanishing. We note here that these Eqs. 14 are often used to predict squeezing in a linearized fluctuation analysis. They are non-classical in the sense that they can describe states without a positive Glauber-Sudarshan P-distribution [35], but correspond to a simple form of linear fluctuation which has a Gaussian quasi-probability distribution.

We now find it useful to introduce combined field quadratures which include both amplified modes of the system, as in two-mode approaches used previously [36]. This is useful here, because unlike the degenerate case, the signal and idler modes exhibit correlated quadrature noise statistics, even though there is no phase sensitivity to the noise if they are treated separately. However, the high degree of cross-correlation found means that combined quadratures can present a high degree of noise suppression, which may be measured by homodyne techniques [37]. We define the two-mode quadratures as

$$x_\pm = \frac{x_1^{(1)} \pm x_2^{(1)}}{\sqrt{2}}, \qquad y_\pm = \frac{y_1^{(1)} \pm y_2^{(1)}}{\sqrt{2}}. \tag{16}$$

Taking the steady state solutions for the pumped field quadratures, we can write the following equations for the linear quantum fluctuations in the new quadratures,

$$dx_+ = \{-Ax_+ + Ex_-\}\,d\tau + \sqrt{\frac{x_{0s}}{2}}\,[dw_{x_1} + dw_{x_2}],$$
$$dx_- = \{-Bx_- + Ex_+\}\,d\tau + \sqrt{\frac{x_{0s}}{2}}\,[dw_{x_1} - dw_{x_2}],$$
$$dy_+ = \{-Cy_+ + Ey_-\}\,d\tau - i\sqrt{\frac{x_{0s}}{2}}\,[dw_{y_1} + dw_{y_2}],$$
$$dy_- = \{-Dy_- + Ey_+\}\,d\tau - i\sqrt{\frac{x_{0s}}{2}}\,[dw_{y_1} - dw_{y_2}], \qquad (17)$$

where we have defined

$$A = 1 - \frac{x_{0s}}{2} + \left(\frac{x_{1s} + x_{2s}}{2}\right)^2$$
$$B = 1 + \frac{x_{0s}}{2} + \left(\frac{x_{1s} - x_{2s}}{2}\right)^2$$
$$C = 1 + \frac{x_{0s}}{2} + \left(\frac{x_{1s} + x_{2s}}{2}\right)^2$$
$$D = 1 - \frac{x_{0s}}{2} + \left(\frac{x_{1s} - x_{2s}}{2}\right)^2,$$
$$E = \left(x_{1s}^2 - x_{2s}^2\right)/4. \qquad (18)$$

These are linear coupled stochastic equations, and we may readily calculate the steady state averages of the first-order corrections and use that to compute the fluctuations in the combined quadratures. These quantities correspond to the squeezed and anti-squeezed combined quadratures obtained in the linearized theory.

## V. CORRELATIONS AND NOISE SPECTRA

In an experimental situation, the noise spectra outside the cavity are generally the quantities of interest. We will therefore proceed to analyze the problem in frequency space, via Fourier decomposition of the fields. The full nonlinear spectra can be found by Fourier transform of the results of stochastic integration of the full positive-P representation equations, which must be performed numerically. However, we will also find it useful to calculate these spectra using our perturbation approach.

We define the Fourier transform as

$$\tilde{f}(\Omega) = \frac{1}{\sqrt{2\pi}} \int_{-\infty}^{+\infty} d\tau\, e^{-i\Omega\tau} f(\tau). \qquad (19)$$

We also need to represent the white noise that drives the stochastic equations by its Fourier transform, $\xi_{x,y}(\Omega)$, where the spectral moments of the stochastic processes are

$$\langle \xi_a(\Omega) \rangle = 0,$$
$$\langle \xi_{a1}(\Omega) \xi_{b2}(\Omega') \rangle = \delta_{ab}\delta(\Omega + \Omega'). \tag{20}$$

The first order stochastic equations for the combined quadratures may now be rewritten in the frequency domain as

$$i\Omega \tilde{x}_+(\Omega) = -A\tilde{x}_+(\Omega) + E\tilde{x}_-(\Omega) + \sqrt{\frac{x_{0s}}{2}}\left[\xi_{x1}(\Omega) + \xi_{x2}(\Omega)\right],$$
$$i\Omega \tilde{x}_-(\Omega) = -B\tilde{x}_-(\Omega) + E\tilde{x}_+(\Omega) + \sqrt{\frac{x_{0s}}{2}}\left[\xi_{x1}(\Omega) - \xi_{x2}(\Omega)\right],$$
$$i\Omega \tilde{y}_+(\Omega) = -C\tilde{y}_+(\Omega) + E\tilde{y}_-(\Omega) - i\sqrt{\frac{x_{0s}}{2}}\left[\xi_{y1}(\Omega) + \xi_{y2}(\Omega)\right],$$
$$i\Omega \tilde{y}_-(\Omega) = -D\tilde{y}_-(\Omega) + E\tilde{y}_+(\Omega) - i\sqrt{\frac{x_{0s}}{2}}\left[\xi_{y1}(\Omega) - \xi_{y2}(\Omega)\right]. \tag{21}$$

Using the above results, we may now calculate the spectra of the squeezed and anti-squeezed field quadratures. These quantities are related in a simple manner to the spectra outside the cavity and, as will be shown below, allow for a measure of entanglement between the two modes.

The contribution from first order perturbation theory is the usual linearized intracavity squeezing result, given in this case by

$$\langle \tilde{y}_+(\Omega) \tilde{y}_+(\Omega') \rangle = \frac{-x_{0s}\left(\Omega^2 + D^2 - E^2\right)\delta(\Omega + \Omega')}{\Omega^2\left(\Omega^2 + C^2 + D^2 + 2E^2\right) + (CD - E^2)^2}, \tag{22}$$

and

$$\langle \tilde{x}_-(\Omega) \tilde{x}_-(\Omega') \rangle = \frac{-x_{0s}\left(\Omega^2 + A^2 - E^2\right)\delta(\Omega + \Omega')}{\Omega^2\left(\Omega^2 + A^2 + B^2 + 2E^2\right) + (AB - E^2)^2}, \tag{23}$$

while the complementary antisqueezed spectra are given by

$$\langle \tilde{x}_+(\Omega) \tilde{x}_+(\Omega') \rangle = \frac{x_{0s}\left(\Omega^2 + B^2 - E^2\right)\delta(\Omega + \Omega')}{\Omega^2\left(\Omega^2 + A^2 + B^2 + 2E^2\right) + (AB - E^2)^2}, \tag{24}$$

and

$$\langle \tilde{y}_-(\Omega) \tilde{y}_-(\Omega') \rangle = \frac{x_{0s}\left(\Omega^2 + C^2 - E^2\right)\delta(\Omega + \Omega')}{\Omega^2\left(\Omega^2 + C^2 + D^2 + 2E^2\right) + (CD - E^2)^2}. \tag{25}$$

The external spectra are obtained in the positive-P representation by the relation

$$S_{ij}^{out}(\Omega)\delta(\Omega + \Omega') = \delta_{ij} + 2\sqrt{\gamma_i^{out}\gamma_j^{out}}\langle \Delta X_i(\Omega) \Delta X_j(\Omega') \rangle_P. \tag{26}$$

Using these results, we find that the external squeezing spectra for the combined quadratures $y_+$ and $x_-$ are different, and given by

$$S_{y_+}^{out}(\Omega) = 1 - \frac{2x_{0s}\left(\Omega^2 + D^2 - E^2\right)}{\Omega^2\left(\Omega^2 + C^2 + D^2 + 2E^2\right) + (CD - E^2)^2} \tag{27}$$

and

$$S_{x_-}^{out}(\Omega) = 1 - \frac{2x_{0s}\left(\Omega^2 + A^2 - E^2\right)}{\Omega^2\left(\Omega^2 + A^2 + B^2 + 2E^2\right) + (AB - E^2)^2}. \tag{28}$$

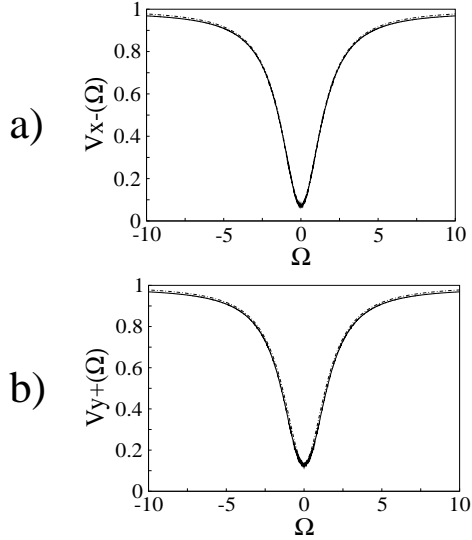

FIG. 2: Dimensionless noise spectra of a) $x_-$ and b) $y_+$, for below threshold operation with $\mu_0 = 0.6$ and $\mu_1 = 0.2$. Frequencies ($\Omega$) are expressed in units of the cavity damping rate for the down converted fields ($\gamma$). Solid lines are the spectra obtained from numerical simulations of the complete stochastic equations in the positive P representation. Dashed lines are the analytical results given by the first order perturbation approach.

These results can now be compared with those found via stochastic integration of the full equations of motion. What we immediately see, as shown in Fig. 2 is that our approximation is a good representation for the noise spectrum of the combined quadratures in the region of small injected signal and pump intensity below the classical threshold ($\mu_1 = 0$ and $\mu_0 = 1$). We also note here that, for smaller values of $\mu_0$ and $\mu_1$, the agreement between the approximate and numerical solutions improves, indicating that the parameter values used in the graphics are at the limit of validity of our approximations.

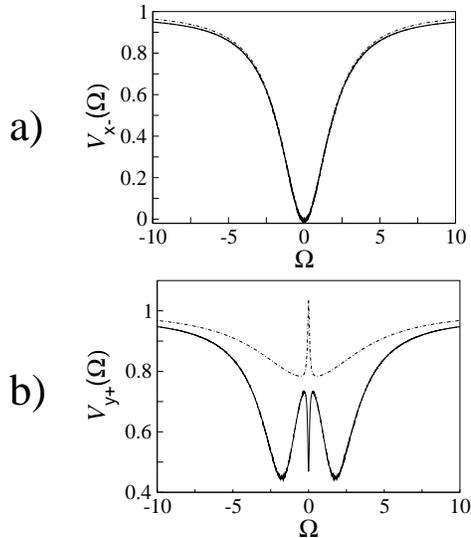

FIG. 3: Dimensionless noise spcetra of a) $x_-$ and b) $y_+$, for above threshold operation with $\mu_0 = 2$ and $\mu_1 = 0.2$. Frequencies ($\Omega$) are expressed in units of the cavity damping rate for the down converted fields ($\gamma$). Solid lines are the spectra obtained from numerical simulations of the complete stochastic equations in the positive P representation. Dashed lines are the analytical results given by the first order perturbation approach.

It is important to stress that to construct the linear noise spectrum in any region of the phase space one has to include the correct steady-state solutions (see equations (9), (10), (11), and (12)) in the particular region. Also, above the classical threshold, where nonlinearities become important, we see that $V_{x_-}(\Omega)$ does not change noticeably, while $V_{y_+}(\Omega)$ is degraded and the first order approximation is no longer valid, as described in Fig. 3. The suitability of a linearized calculation for $V_{x_-}(\Omega)$, but not for $V_{y_+}(\Omega)$, is not surprising if we recall that, for a sufficiently injected OPO, $\langle x_- \rangle \gg \sqrt{\langle (\delta x_-)^2 \rangle}$. In this case, the fluctuations are much smaller than the steady state average and a linearized approach applies. However, $\langle y_+ \rangle = 0$ so that nonlinear contributions of the fluctuations cannot be neglected. For the case of triple resonance operation considered here the stochastic equations for $x_-$ and $y_+$ are decoupled, and each variable admits an independent treatment.

From numerical integration of the complete stochastic equations we find, similarly to the phase variance described previously in Ref. [7], that $V_{y_+}(\Omega)$ becomes bifurcated and shows less absolute squeezing above the threshold. Even though our quadratures are not exactly comparable to the intensity difference and phase sum variables considered in that work, it is noticeable that they display similar behaviors.

## VI. QUANTUM ENTANGLEMENT AND THE EPR PARADOX

Entanglement is a property found in quantum mechanical systems and exists when the combined density matrix cannot be factorized into a product of the density matrices for the component subsystems. An entanglement criterion for continuous variables has been developed by Duan *et al.* [38] which are based on the inseparability of the system density matrix. We will briefly outline these criteria here and then apply them to our system. Following the treatment of Ref. [32], and noting the quadrature normalization used above (Eq. 16), entanglement is guaranteed provided that the sum of the variances of these quadratures is less than 2.

A direct and feasible demonstration of the EPR correlations with continuous variables was first suggested by Reid [39]. The proposal was for an optical demonstration of the paradox via quadrature phase amplitudes, using nondegenerate parametric amplification. The quadrature phase amplitudes used have the same mathematical properties as the position and momentum originally used by EPR and, even though the correlations between these are not perfect, they are still entangled sufficiently to allow for an inferred violation of the uncertainty principle. An experimental demonstration of Reid's proposal by Ou *et al.* [40] soon followed, showing a clear agreement with quantum theory.

Let us calculate the correlations necessary for a demonstration of the EPR correlations. Following the approach of Reid [39], we assume that a measurement of the $x_1$ quadrature, for example, will allow us to infer, with some error, the value of the $x_2$ quadrature, and similarly for the $y_j$ quadratures. This allows us to make linear estimates of the quadrature variances, which are then minimized to give the inferred output variances,

$$\begin{aligned}
V^{\inf}[x_1(\Omega)] &= S^{out}_{x_1}(\Omega) - \frac{|S^{out}_{x_1,x_2}(\Omega)|^2}{S^{out}_{x_2}(\Omega)}, \\
V^{\inf}[y_1(\Omega)] &= S^{out}_{y_1}(\Omega) - \frac{|S^{out}_{y_1,y_2}(\Omega)|^2}{S^{out}_{y_2}(\Omega)}.
\end{aligned} \quad (29)$$

The inferred variances for the $j = 2$ quadratures are simply found by swapping the indices 1 and 2. As the $\hat{X}_j$ and $\hat{Y}_j$ operators do not commute, the products of the variances obey a Heisenberg uncertainty relation, with $V(x_j)V(y_j) \geq 1$. Hence we find a demonstration of the EPR correlations whenever

$$V^{\text{inf}}\left[x_j(\Omega)\right] V^{\text{inf}}\left[y_j(\Omega)\right] \leq 1. \tag{30}$$

To first order in our perturbation approach, we may calculate all the correlations necessary to show the paradox. Again working with the frequency components and using the results of Eq. 21, we find the following correlations

$$\langle \tilde{x}_+(\Omega)\tilde{x}_-(\Omega')\rangle + \langle \tilde{x}_-(\Omega)\tilde{x}_+(\Omega')\rangle = \frac{2x_{0s}E(B-A)\delta(\Omega+\Omega')}{\Omega^2(\Omega^2+A^2+B^2+2E^2)+(AB-E^2)^2} \tag{31}$$

and

$$\langle \tilde{y}_+(\Omega)\tilde{y}_-(\Omega')\rangle + \langle \tilde{y}_-(\Omega)\tilde{y}_+(\Omega')\rangle = \frac{-2x_{0s}E(D-C)\delta(\Omega+\Omega')}{\Omega^2(\Omega^2+C^2+D^2+2E^2)+(CD-E^2)^2}, \tag{32}$$

which allows us to write the inferred variances as

$$\begin{aligned}
V^{inf}\left[y_1(\Omega)\right] &= \frac{1}{2}\left\{\left[S^{out}_{x_+}(\Omega) + S^{out}_{y_+,y_-}(\Omega) + S^{out}_{y_-,y_+}(\Omega) + S^{out}_{y_-}(\Omega)\right]\right.\\
&\quad \left.-\frac{|S^{out}_{y_+}(\Omega) - S^{out}_{y_+,y_-}(\Omega) + S^{out}_{y_-,y_+}(\Omega) - S^{out}_{y_-}(\Omega)|^2}{S^{out}_{y_+}(\Omega) - S^{out}_{y_+,y_-}(\Omega) - S^{out}_{y_-,y_+}(\Omega) + S^{out}_{y_-}(\Omega)}\right\},\\
V^{inf}\left[x_1(\Omega)\right] &= \frac{1}{2}\left\{\left[S^{out}_{x_+}(\Omega) + S^{out}_{x_+,x_-}(\Omega) + S^{out}_{x_-,y_+}(\Omega) + S^{out}_{x_-}(\Omega)\right]\right.\\
&\quad \left.-\frac{|S^{out}_{x_+}(\Omega) - S^{out}_{x_+,x_-}(\Omega) + S^{out}_{x_-,x_+}(\Omega) - S^{out}_{x_-}(\Omega)|^2}{S^{out}_{x_+}(\Omega) - S^{out}_{x_+,x_-}(\Omega) - S^{out}_{x_-,x_+}(\Omega) + S^{out}_{x_-}(\Omega)}\right\},
\end{aligned} \tag{33}$$

and similarly for the second mode. Using the solutions previously obtained, and following some algebra, we can calculate the product $V^{inf}[y_1(\Omega)]V^{inf}[x_1(\Omega)]$.

In order to provide a complete description of the entanglement properties of the injected OPO, we will consider both the EPR criterion established in terms of inferred variances and the Duan criterion. We shall develop our analysis both in frequency and quadrature domain. Arbitrary quadratures $X(\theta)$ and $Y(\theta)$ are given by a simple rotation transformation according to

$$\begin{aligned}
X(\theta) &= X\cos\theta + Y\sin\theta, \\
Y(\theta) &= -X\sin\theta + Y\cos\theta.
\end{aligned} \tag{34}$$

The noise and correlation spectra can then be easily calculated for any value of $\theta$.

In Fig.4 we show the noise sum and the product of inferred variances for an injected OPO ($\mu_1 = 0.2$) operating below threshold ($\mu_0 = 0.6$). The limiting values for violation of classical behavior (1 for the EPR criterion and 2 for the Duan criterion) are also indicated. Since the linearized approach is valid in this regime, we used our analytical result in these curves. For $\theta = 0$ violation is obtained in most of the frequency range according to both criteria (Fig.4a). In Fig.4b we plot the zero frequency evaluation of the criteria as a function of $\theta$. In this case, the EPR criterion is more restrictive and it is interesting to observe that there exists a range of quadratures for which only one of the criteria is satisfied.

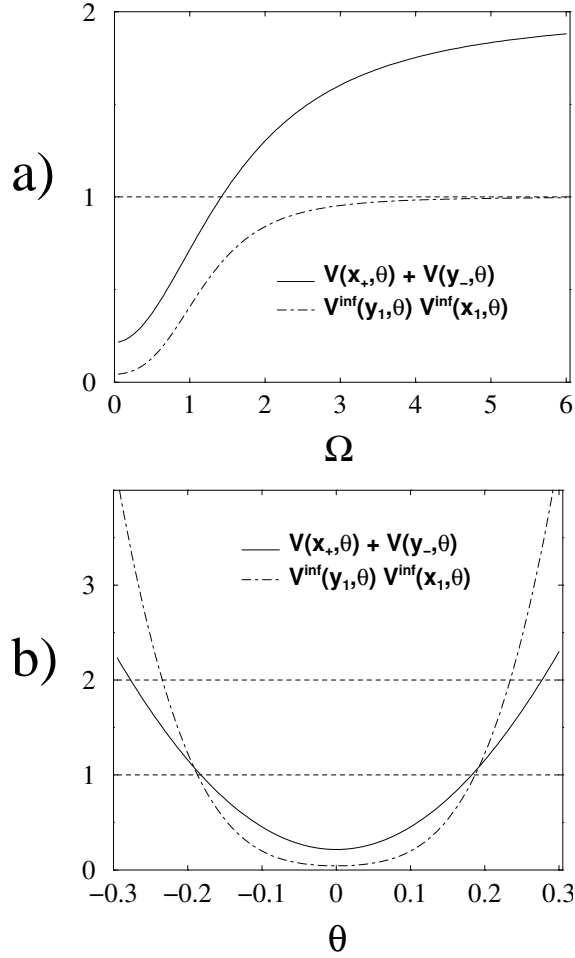

FIG. 4: a) Dimensionless spectra of the Duan and EPR criteria, evaluated below threshold operation ($\mu_0 = 0.6$ and $\mu_1 = 0.2$) for $\theta = 0\,rad$, obtained with first order perturbation. Frequencies ($\Omega$) are expressed in units of the cavity damping rate ($\gamma$) for the down converted fields. b) Dimensionless spectra of the Duan and EPR criteria at zero frequency as a function of $\theta$ (expressed in radians) under the same conditions of a).

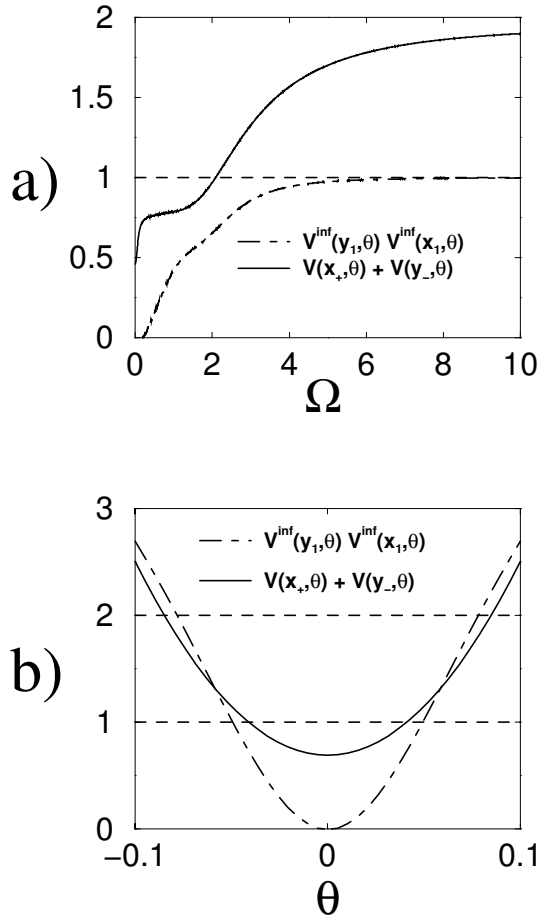

FIG. 5: a) Dimensionless spectra of the Duan and EPR criteria, evaluated above threshold operation ($\mu_0 = 2$ and $\mu_1 = 0.2$) for $\theta = 0\,rad$, obtained with numerical simulations of the Eqs.(2). Frequencies ($\Omega$) are expressed in units of the cavity damping rate ($\gamma$) for the down converted fields. b) Dimensionless spectra of the Duan and EPR criteria at zero frequency as a function of $\theta$ (expressed in radians) under the same conditions of a).

In Fig.5 we show the noise sum and the product of inferred variances for an injected OPO ($\mu_1 = 0.2$) operating above threshold ($\mu_0 = 2$). Since the linearized approach does not apply to this condition, we used only numerical results in these curves. For $\theta = 0$ violation is obtained in most of the frequency range according to both criteria (Fig.5a). In Fig.5b we plot the zero frequency evaluation of the criteria as a function of $\theta$. Again, the EPR criterion is more restrictive. Moreover, violation of classical behavior occurs in a quadrature range narrower than the one in Fig.4b. In both cases this quadrature range is quite narrow, which means that a considerable control of the local oscillator is required in order to render this effect experimentally observable.

In Fig.6 we plot the two criteria below threshold ($\mu_0 = 0.6$) for different injection levels. It allows for a comparison between the injected and the uninjected cases. Let us consider the intensities of the fields which are involved in this demonstration. The quadrature first-order steady state values are as given previously, in Eqs. 9 and 10, from these we can calculate the normalized intensities of the down-converted fields $I_j = x_{js}^2$ ($j = 1, 2$). For $\mu_1 = 0.1$ we see little difference from the uninjected result, but with intensities of $I_{1s} = 0.088$ and $I_{2s} = 0.028$. Considering the case of $\mu_1 = 0.2$, we find that $I_{1s} = 0.28$ and $I_{2s} = 0.072$, while a significant violation of the classical limit is predicted. Although these numbers in themselves do not seem very large, we point out that for typical experimental setups, this would be a truly intense field.

For example, let us consider an OPO with one input-output mirror with transmission coefficient $T_0 = 2\gamma_0 \tau$ for the pump field ($R \sim 100\%$ for the down-converted fields) and another input-output mirror with a common transmission coefficient $T = 2\gamma\tau$ for signal and idler ($R \sim 100\%$ for the pump field), where $\tau$ is the cavity round trip time. It is convenient to express the powers involved in the OPO operation in terms of the threshold power $P_{th}$. The pump power is simply $P_0 = \mu_0^2 P_{th}$, the injection power is $P_1 = 2\mu_1^2 P_{th}(\omega_1/\omega_0)$ and the output powers of the down-converted fields are $P_j^{out} = 2x_{js}^2 P_{th}(\omega_j/\omega_0)$. If we consider $P_{th} = 20mW$, which is a reasonable experimental value, and $\omega_1 \sim \omega_2 \sim \omega_0/2$, then our numerical example with $\mu_0 = 0.6$ and $\mu_1 = 0.2$ gives $P_0 = 7.2mW$, $P_1 = 1.6mW$, $P_1^{out} = 5.6mW$ and $P_2^{out} = 1.4mW$.

## VII. CONCLUSIONS

We have shown how an injected signal in one of the lower frequency modes of the non-degenerate optical parametric oscillator can be used to increase the intensity of both low-frequency modes below the normal oscillation threshold, while not markedly degrading the quantum correlations. Using a perturbation expansion of the full quantum equations of motion, we have shown the effects of this injected signal on the quantum properties of the system. Our analysis shows that a high degree of entanglement and a good demonstration of the EPR correlations are possible with reasonable injection values. Moreover, the amplified fields in this case are now macroscopic, rather than being composed essentially of fluctuations as in previous experimental demonstrations of these qualities. In conclusion, we predict that this device may be useful for the production of continuous variable entangled and EPR states with macroscopic intensities.

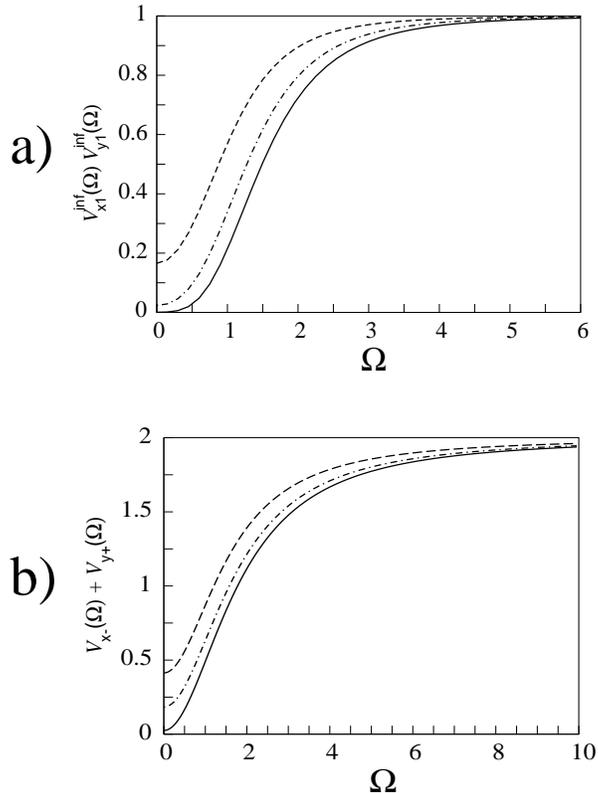

FIG. 6: Dimensionless spectra of the EPR (a) and Duan (b) criteria for $\theta = 0\,rad$ and below threshold operation ($\mu_0 = 0.6$), obtained with first order perturbation. $\mu_1 = 0$ (solid line), $\mu_1 = 0.1$ (dot-dashed line) and $\mu_1 = 0.2$ (dashed line). Frequencies ($\Omega$) are expressed in units of the cavity damping rate for the down converted fields ($\gamma$).

**Acknowledgments**

We acknowledge financial support from the Brazilian agencies CNPq (Conselho Nacional de Desenvolvimento Científico e Tecnológico), CAPES (Coordenadoria de Aperfeiçoamento de Pessoal de Nível Superior) and the Australian Research Council.

**APPENDIX A: APPROXIMATE STEADY STATE SOLUTIONS**

To solve the steady state equations for $x_{0s}$, we may write a fifth order product,

$$P_5(x) = (2\mu_0 - x)\left(1 - \frac{1}{4}x^2\right)^2, \tag{A1}$$

the solutions being found as the intersections of $P_5(x)$ with the straight line $f(x) = 2\mu_1^2 x$, that is, as the real roots of the polynomial equation $P_5(x) - f(x) = 0$. Setting $\mu_1 = 0$ leads to the usual solutions for the non-injected OPO: $x_{0s} = 2\mu_0$, which is the only stable solution below threshold ($\mu_0 < 1$), and $x_{0s} = \pm 2$, which are the two stable solutions above threshold.

In order to find approximate analytical solutions for $x_{0s}$, we shall adopt the strategy of dividing the operation into three different regions, namely below, above and at threshold. In each regime we shall make a suitable approximation for $P_5(x)$, always assuming a weak injection ($\mu_1 \ll 1$). For the below threshold regime, we can make a linear approximation for $P_5(x)$ around $2\mu_0$ by setting $x = 2\mu_0 + \epsilon$ and keeping terms up to first order in $\epsilon$. Above threshold ($\mu_0 > 1$), there may be up to three steady state solutions for $x_{0s}$. However, as we will show in appendix B, only the solution for which $x_{0s} < 2$ is stable. Moreover, the shape of $P_5(x)$ for above threshold operation suggests that a quadratic approximation can be employed around $x = 2$. Therefore we will set $x = 2 + \epsilon$ and keep terms up to second order in $\epsilon$. Finally, the shape of $P_5(x)$ at threshold suggests a cubic approximation around $x = 2$. However, as we will show, we found that a quartic approximation gives a simpler result.

In Fig. 7 we show a graphical representation of the steady state solution for $x_{0s}$ in each regime. In all cases, the full solution is found as the intersection of the full line representing $P_5(x)$ with the long-dashed line of positive slope representing $f(x) = 2\mu_1^2 x$. The linear approximation corresponds to approximating $P_5(x)$ by a tangent straight line and finding its intersection with $f(x)$. As can be seen from the graph in Fig. 7a, the two intersections are in close proximity. This approximation gives the following solution for the steady-state amplitude of the pump field quadratures below threshold:

$$x_{0s} = 2\mu_0 \left[1 - \frac{2\mu_1^2}{(1-\mu_0^2)^2 + 2\mu_1^2}\right] . \tag{A2}$$

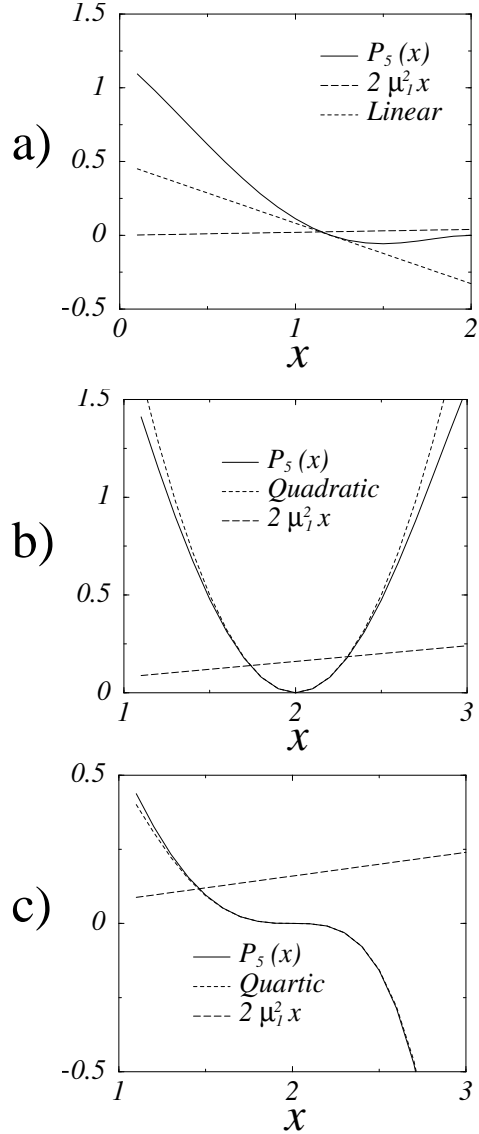

FIG. 7: Graphic representation of the steady state solution for the dimensionless amplitude quadrature $x_{0s}$. a) Below threshold ($\mu_0 = 0.6$). A linear approximation (short-dashed) is used for the dimensionless polinomial $P_5(x)$. b) Above threshold ($\mu_0 = 2$). A quadratic approximation (short-dashed) is used for $P_5(x)$. c) On threshold ($\mu_0 = 1$). A quartic approximation (short-dashed) is used for $P_5(x)$. In all figures we considered $\mu_1 = 0.2$, the solid lines are the full evaluation of $P_5(x)$, and the long-dashed lines represent the dimensionless quantity $2\,\mu_1^2\,x$.

To obtain solutions in the second, above threshold region, we approximate $P_5(x)$ by a tangent parabola as shown in Fig. 7b. The two solutions for $x_{0s}$ around $x = 2$ are then obtained as the intersections of $f(x)$ with this parabola, the smaller of these being the only stable one. As can be seen from the figure, these approximate solutions are almost indistinguishable from the full solutions, found as the intersections of $P_5(x)$ with $f(x)$. The stable approximate solution found in this case is then

$$x_{0s} = 2 - \frac{\mu_1^2}{2(\mu_0 - 1)} \left[\sqrt{1 + \frac{8(\mu_0 - 1)}{\mu_1^2}} - 1\right], \tag{A3}$$

Finally, the quartic approximation, valid at threshold ($\mu_0 = 1$), gives a quite simple solution

$$x_{0s} = 2 - (2\mu_1)^{2/3}. \tag{A4}$$

This solution for $x_{0s}$ is shown in Fig. 7c, as the intersection of the quartic approximation and $f(x)$. As can be seen, the fourth-order polynomial approximates $P_5(x)$ very closely in this region.

### APPENDIX B: STABILITY ANALYSIS OF THE STEADY STATE SOLUTIONS

We now present a linear stability analysis of the steady state solutions of section III. Let us consider small deviations of the dynamical variables from their steady state average values as follows:

$$\delta x_j = x_{js} - \langle x_j \rangle$$
$$\delta y_j = y_{js} - \langle y_j \rangle , \tag{B1}$$

with $j = 0, 1, 2$. From the average of Eqs.7, we can write down the equations of motion for $\delta x_j$ and $\delta y_j$. The linearized equations of motion are then obtained by keeping terms up to first order in these deviations. Defining the column vectors

$$\delta \vec{x} = \begin{bmatrix} \delta x_0 \\ \delta x_1 \\ \delta x_2 \end{bmatrix} \quad \text{and} \quad \delta \vec{y} = \begin{bmatrix} \delta y_0 \\ \delta y_1 \\ \delta y_2 \end{bmatrix} , \tag{B2}$$

we can write the linearized equations in a compact form:

$$\delta\dot{\vec{x}} = \mathbf{M_x}\,\delta\vec{x} \qquad \text{and} \qquad \delta\dot{\vec{y}} = \mathbf{M_y}\,\delta\vec{y}\,, \tag{B3}$$

where

$$\mathbf{M_x} = \begin{bmatrix} -\gamma_r & -\gamma_r\,x_{2s} & -\gamma_r\,x_{1s} \\ x_{2s}/2 & -1 & x_{0s}/2 \\ x_{1s}/2 & x_{0s}/2 & -1 \end{bmatrix} \qquad \text{and} \qquad \mathbf{M_y} = \begin{bmatrix} -\gamma_r & -\gamma_r\,x_{2s} & -\gamma_r\,x_{1s} \\ x_{2s}/2 & -1 & -x_{0s}/2 \\ x_{1s}/2 & -x_{0s}/2 & -1 \end{bmatrix}. \tag{B4}$$

These steady state solutions are stable if all eigenvalues of $\mathbf{M_x}$ and $\mathbf{M_y}$ have negative real parts. We therefore arrive at the following secular equations for the eigenvalues $\lambda_x$ and $\lambda_y$:

$$\begin{aligned} \lambda_x^3 + c_1\lambda_x^2 + c_2\lambda_x + c_3 &= 0 \\ \lambda_y^3 + d_1\lambda_y^2 + d_2\lambda_y + d_3 &= 0\,, \end{aligned} \tag{B5}$$

where

$$\begin{aligned} c_1 &= d_1 = 2 + \gamma_r\,, \\ c_2 &= d_2 = 1 - x_{0s}^2/4 + 2\gamma_r + \frac{\gamma_r}{2}\left(x_{1s}^2 + x_{2s}^2\right), \\ c_3 &= \gamma_r\left[1 - x_{0s}^2/4 + (x_{1s}^2 + x_{2s}^2)/2 + x_{0s}\,x_{1s}\,x_{2s}/2\right], \\ d_3 &= \gamma_r\left[1 - x_{0s}^2/4 + (x_{1s}^2 + x_{2s}^2)/2 - x_{0s}\,x_{1s}\,x_{2s}/2\right]. \end{aligned} \tag{B6}$$

According to the Hurwitz criterion [41], all eigenvalues have negative real parts if and only if the coefficients $c_1$, $c_2$, $c_3$, $d_1$, $d_2$, $d_3$ are all positive and the Hurwitz determinants $H_1 = c_1$, $H_2 = c_1\,c_2 - c_3$, $H_3 = c_3\,H_2$, $G_1 = d_1$, $G_2 = d_1\,d_2 - d_3$, and $G_3 = d_3\,G_2$, are also positive.

At this point, it is useful to write all the conditions in terms of the system parameters and $x_{0s}$. From the steady state given by Eqs. 8 we have

$$x_{1s} = \frac{8\,\mu_1}{4 - x_{0s}^2}\,, \tag{B7}$$

and

$$x_{2s} = \frac{4\mu_1\,x_{0s}}{4 - x_{0s}^2}\,. \tag{B8}$$

After some straightforward algebra and careful inspection of the above conditions, one can reduce the whole Hurwitz criterion to a single condition:

$$(1 - x_{0s}^2/4)\left[2\mu_1^2 + (1 - x_{0s}^2/4)^2\right] > 0\,. \tag{B9}$$

Therefore, it can be clearly seen that any steady state solution with $x_{0s} > 2$ is unstable. In practice, this means that all the physical solutions we find with injected signal are stable, in contrast to the uninjected case where there is an instability at the oscillation threshold.

---